\documentclass[fleqn,twoside]{article}
\usepackage{espcrc2}

\def\sym#1{{\mathcal #1}}

\newcommand{\href}[2]{{\texttt{#2}}}%
\newcommand{\dhref}[2]{}
\def\Wloop{W(\{\tau_{j\kappa}\},\{Q^{(j)}\})}
\def\hpeprint#1{{\texttt #1}}
\def\hpspires#1{}

\let\phi=\varphi
\let\theta=\vartheta
\let\epsilon=\varepsilon

\makeatletter
\newfont{\@aidxte}{cmsy10}
\newfont{\@aidxel}{cmsy10 scaled 1095}
\newfont{\@aidxtw}{cmsy10 scaled 1200}
\newlength\@aidxtexvi
\newlength\@aidxtexvii
\newlength\@aidxelxvi
\newlength\@aidxelxvii
\newlength\@aidxtwxvi
\newlength\@aidxtwxvii
\newcommand{\alignidx}[1]{%
  \@aidxtexvi=\fontdimen16\@aidxte
  \@aidxtexvii=\fontdimen17\@aidxte
  \@aidxelxvi=\fontdimen16\@aidxel
  \@aidxelxvii=\fontdimen17\@aidxel
  \@aidxtwxvi=\fontdimen16\@aidxtw
  \@aidxtwxvii=\fontdimen17\@aidxtw
    {\mbox{$%
    \fontdimen16\@aidxte=2.9pt
    \fontdimen17\@aidxte=2.9pt
    \fontdimen16\@aidxel=3.1pt
    \fontdimen17\@aidxel=3.1pt
    \fontdimen16\@aidxtw=3.3pt
    \fontdimen17\@aidxtw=3.3pt
    #1$}}%
    \fontdimen16\@aidxte=\@aidxtexvi
    \fontdimen17\@aidxte=\@aidxtexvii
    \fontdimen16\@aidxel=\@aidxelxvi
    \fontdimen17\@aidxel=\@aidxelxvii
    \fontdimen16\@aidxtw=\@aidxtwxvi
    \fontdimen17\@aidxtw=\@aidxtwxvii}

\makeatother

\title{The dual of non-Abelian Lattice Gauge Theory}
\author{Hendryk Pfeiffer\address{Department of Applied Mathematics and Theoretical Physics,
  Cambridge CB3 0WA, UK} and Robert Oeckl\address{Centre de Physique
  Th\'eorique, Campus de Luminy, 13288 Marseille, France}}

\begin{document}

%
\begin{abstract}
%

  Non-Abelian Lattice Gauge Theory in Euclidean space-time of
  dimension $d\geq 2$ whose gauge group is any compact Lie group is
  related to a Spin Foam Model by an exact strong-weak duality
  transformation. The group degrees of freedom are integrated out and
  replaced by combinatorial expressions involving irreducible
  representations and intertwiners of the gauge group.  This
  transformation is available for the partition function, for the
  expectation value of observables (spin networks), and for the
  correlator of centre monopoles which is a ratio of partition
  functions in the original model and an ordinary expectation value in
  the dual formulation.
\end{abstract}

\maketitle

%
\section{Introduction}
%

Exact duality transformations in Statistical Mechanics have been known
since the discovery of the self-duality of the $2$-dimensional Ising
model. This transformation was generalized to a wider class of Abelian
lattice models including Abelian Lattice Gauge Theory
(LGT)~\cite{Sa80,Pe78}. In $d=4$, this transformation for pure $U(1)$
LGT can be understood as the lattice version of electric-magnetic
duality. It proved to be very powerful in the analytical~\cite{FrMa86}
and numerical~\cite{JeNe99} study of magnetic monopoles and their
role in confinement.

We present a generalization of this transformation to non-Abelian LGT
in $d\geq 2$ with a compact Lie group $G$ as the gauge group. A
similar transformation has been known for the partition function in
the special case of $SU(2)$ in $d=3$~\cite{AnCh93} for which
numerical algorithms are being studied~\cite{HaSh01}. For the duality
transformation, all group integrals are solved, and the group degrees
of freedom are replaced by combinatorial expressions involving
irreducible representations and intertwiners of the gauge
group. Mathematically, this transformation can be understood as an
explicit calculation exploiting Tannaka--Kre\v\i n duality which
provides a one-to-one correspondence between compact Lie groups and
suitable tensor categories which arise as the categories of
finite-dimensional representations of these Lie groups.

In the duality transformation, the Boltzmann weight $\exp(-s(g_P))$ is
character expanded, generalizing the Fourier expansion of
the $U(1)$ case. For the standard local actions $s(g_P)$, such as
Wilson's action or the heat kernel action, this ensures that the
duality transformation still provides the strong-weak relation
as for the Abelian case. The dual expressions for
expectation values of generic observables can thus be used to generate
their strong coupling expansion.

The dual model is a Spin Foam Model very similar to those models which
are currently studied in the context of non-perturbative quantum
gravity~\cite{Ba99,Or01} and to the state sums used in order to define
topological field theory and invariants~\cite{TuVi92,CrKa97}. The
duality transformation has been described in detail in~\cite{OePf01}
for hypercubic lattices and Lie groups. For an extension to generic
triangulations and to quantum groups, see~\cite{Oe01,Pf01}.

As for the Abelian special case, we find that expectation values
are mapped under the transformation to ratios of partition functions
and vice versa. In particular, there is an observable of the dual
model whose expectation value agrees with the correlator of centre
monopoles and vortices.

%
\section{The duality transformation}
%

\subsection{The partition function}

The partition function of pure non-Abelian LGT on a $d$-dimensional
hypercubic lattice reads
\begin{equation}
  Z = \Bigl(\prod_{i,\mu}\int\limits_G\,dg_{i\mu}\Bigr)\,
      \prod_{i,\mu<\nu}f(\alignidx{g_{i\mu}g_{i+\hat\mu\nu}
            g_{i+\hat\nu\mu}^{-1}g_{i\nu}^{-1}}). 
\end{equation}
Here the gauge group $G$ is any compact Lie group, $i$ denotes the
lattice points and $i+\hat\mu$ the nearest neighbour of $i$ in the
$\mu$-direction. The duality transformation for the partition function
involves the expansion of the Boltzmann weight $f(g)=\exp(-s(g))$ into
characters,
\begin{equation}
  f(g)=\sum_{\rho}\hat f_\rho\chi^{(\rho)}(g).
\end{equation}
Here the action $s(g)$ is a suitable class function of $G$. Solving
all group integrals, one obtains the dual form of the partition
function,
\begin{eqnarray}
\label{eq_dualpartition}
  Z&=&\Bigl(\prod_{i,\mu<\nu}\sum_{\rho_{i\mu\nu}}\Bigr)\,
    \Bigl(\prod_{i,\mu}\sum_{P^{(i\mu)}\in\sym{P}_{i\mu}}\Bigr)\nonumber\\
   &\times& \Bigl(\prod_{i,\mu<\nu}\hat 
       f_{\rho_{i\mu\nu}}\Bigr)\,\Bigl(\prod_{i}\,C(i)\Bigr).
\end{eqnarray}
Here the dual path integral is a sum over colourings of the plaquettes
with irreducible representations $\rho_{i\mu\nu}$ and a sum over a all
colourings of the links with elements of a basis $\sym{P}_{i\mu}$ of
$G$-invariant orthogonal projectors onto the trivial component in the
complete decomposition of the tensor product
\begin{displaymath}
  \underbrace{
    (\rho_{i-\hat\lambda,\lambda,\mu}\otimes\rho_{i\lambda\mu}^\ast)\otimes\cdots}_{
    \lambda\in\{1,\ldots,\mu-1\}}\otimes
  \underbrace{
    (\rho_{i-\hat\nu,\mu,\nu}^\ast\otimes\rho_{i\mu\nu})\otimes\cdots}_{
    \nu\in\{\mu+1,\ldots,d\}}.
\end{displaymath}
The weights of the dual partition function are the dual Boltzmann
weight $\hat f_{\rho_{i\mu\nu}}$,
\begin{equation}
  \hat f_\rho =d_\rho\,\int\limits_G\,
  \overline{\chi^{(\rho)}(g)}\,f(g)\,dg,
\end{equation}
for each plaquette, and a group invariant
\begin{eqnarray}
  &&C(i)=\\
  &&\Bigl(\prod_{\mu<\nu}
           \sum_{p_{i-\hat\mu-\hat\nu,\mu,\nu}=1}^{d_{\rho_{i-\hat\mu-\hat\nu,\mu,\nu}}}
           \sum_{q_{i-\hat\nu,\mu,\nu}=1}^{d_{\rho_{i-\hat\nu,\mu,\nu}}}
           \sum_{m_{i-\hat\mu,\mu,\nu}=1}^{d_{\rho_{i-\hat\mu,\mu,\nu}}}
           \sum_{n_{i\mu\nu}=1}^{d_{\rho_{i\mu\nu}}}\Bigr)\nonumber
\end{eqnarray}
\begin{displaymath}
  \times\prod_{\mu=1}^d
     P^{(i\mu)}_{(\underbrace{\scriptstyle m_{i-\hat\lambda,\lambda,\mu}n_{i\lambda\mu}\ldots}_{
       \lambda\in\{1,\ldots,\mu-1\}})
        (\underbrace{\scriptstyle q_{i-\hat\nu,\mu,\nu}n_{i\mu\nu}\ldots}_{
       \nu\in\{\mu+1,\ldots,d\}})}
\end{displaymath}
\begin{displaymath}
  P^{(i-\hat\mu,\mu)}_{(\underbrace{\scriptstyle p_{i-\hat\mu-\hat\lambda,\lambda,\mu}
                                                  q_{i-\hat\mu,\lambda,\mu}\ldots}_{
       \lambda\in\{1,\ldots,\mu-1\}})
        (\underbrace{\scriptstyle p_{i-\hat\mu-\hat\nu,\mu,\nu}m_{i-\hat\mu,\mu,\nu}\ldots}_{
       \nu\in\{\mu+1,\ldots,d\}})},
\end{displaymath}
for each lattice point. Here $d_\rho$ denotes the dimension of the
representation $\rho$. 

\subsection{Gauge invariant observables}

The generic gauge invariant observable of LGT is the spin network
\begin{eqnarray}
  &&\Wloop =\\
  && \Bigl(\prod_{j,\kappa}\sum_{a_{j\kappa},b_{j\kappa}}\Bigr)\,
  \Bigl(\prod_{j,\kappa}
        t_{a_{j\kappa}b_{j\kappa}}^{(\tau_{j\kappa})}(g_{j\kappa})\Bigr)\nonumber\\
  &\times&\Bigl(\prod_{j}
    Q^{(j)}_{(b_{j-\hat1,1}\ldots b_{j-\hat d,d}),(a_{j1}\ldots a_{jd})}\Bigr).\nonumber
\end{eqnarray}
Here $\tau_{j\kappa}$ denotes a colouring of links with irreducible
representations and
\begin{equation}
  Q^{(j)}\colon\bigotimes_{\mu=1}^d\tau_{j-\hat\mu,\mu}\to
    \bigotimes_{\mu=1}^d\tau_{j\mu},
\end{equation}
a colouring of the lattice points with compatible intertwiners. The
matrix elements of the representation $\rho$ are denoted by
$t_{ab}^{(\rho)}(g)$.

The dual expression for the expectation value $\left<\Wloop\right>$ is
given by the same expression as~(\ref{eq_dualpartition}) except that
the sum is over projectors in the decomposition of
\begin{eqnarray}
  &&\underbrace{
    (\rho_{i-\hat\lambda,\lambda,\mu}\otimes\rho_{i\lambda\mu}^\ast)\otimes\cdots}_{
    \lambda\in\{1,\ldots,\mu-1\}}\nonumber\\
  &&\otimes\underbrace{
    (\rho_{i-\hat\nu,\mu,\nu}^\ast\otimes\rho_{i\mu\nu})\otimes\cdots}_{
    \nu\in\{\mu+1,\ldots,d\}}\otimes\tau_{i\mu},
\end{eqnarray}
and the weights per point are given by
\begin{eqnarray}
  &&D(i)=\\
  &&\Bigl(\prod_{\mu}\sum_{a_{i\mu}=1}^{d_{\tau_{i\mu}}}
       \sum_{b_{i-\hat\mu,\mu}=1}^{d_{\tau_{i-\hat\mu,\mu}}}\Bigr)\,
    Q^{(i)}_{(b_{i-\hat1,1}\ldots b_{i-\hat d,d}),(a_{i1}\ldots a_{id})}\nonumber\\
  &&\Bigl(\prod_{\mu<\nu}
           \sum_{p_{i-\hat\mu-\hat\nu,\mu,\nu}=1}^{d_{\rho_{i-\hat\mu-\hat\nu,\mu,\nu}}}
           \sum_{q_{i-\hat\nu,\mu,\nu}=1}^{d_{\rho_{i-\hat\nu,\mu,\nu}}}
           \sum_{m_{i-\hat\mu,\mu,\nu}=1}^{d_{\rho_{i-\hat\mu,\mu,\nu}}}
           \sum_{n_{i\mu\nu}=1}^{d_{\rho_{i\mu\nu}}}\Bigr)\nonumber
\end{eqnarray}
\begin{displaymath}
  \times\prod_{\mu=1}^d
     P^{(i\mu)}_{(\underbrace{\scriptstyle m_{i-\hat\lambda,\lambda,\mu}n_{i\lambda\mu}\ldots}_{
       \lambda\in\{1,\ldots,\mu-1\}})
        (\underbrace{\scriptstyle q_{i-\hat\nu,\mu,\nu}n_{i\mu\nu}\ldots}_{
       \nu\in\{\mu+1,\ldots,d\}}) a_{i\mu}}
\end{displaymath}
\begin{displaymath}
  P^{(i-\hat\mu,\mu)}_{(\underbrace{\scriptstyle p_{i-\hat\mu-\hat\lambda,\lambda,\mu}
                                                  q_{i-\hat\mu,\lambda,\mu}\ldots}_{
       \lambda\in\{1,\ldots,\mu-1\}})
        (\underbrace{\scriptstyle p_{i-\hat\mu-\hat\nu,\mu,\nu}m_{i-\hat\mu,\mu,\nu}\ldots}_{
       \nu\in\{\mu+1,\ldots,d\}}) b_{i-\hat\mu,\mu}}.
\end{displaymath}
The normalized expectation value $\left<\Wloop\right>$ of the spin
network is therefore mapped to a ratio of partition functions in the
dual model.

\subsection{Monopole and vortex correlators}

A similar transformation is available for the correlator of centre
monopoles and vortices. It is given by a ratio of partition functions
$Z(X)/Z$,
\begin{eqnarray}
  Z(X) &=& \Bigl(\prod_{i,\mu}\int\limits_G\,dg_{i\mu}\Bigr)\nonumber\\
      &\times&\prod_{i,\mu<\nu}
       f(\alignidx{g_{i\mu}g_{i+\hat\mu\nu}
        g_{i+\hat\nu\mu}^{-1}g_{i\nu}^{-1}X_{i\mu\nu}}),
\end{eqnarray}
where $X_{i\mu\nu}$ associates an element of the centre $Z(G)$ to each
plaquette. Its dual expression has the form of an expectation value
under the dual path integral~(\ref{eq_dualpartition}),
\begin{equation}
  Z(X) = \left<\prod_{i,\mu<\nu}
         {\tilde t}^{(\rho_{i\mu\nu})}(X_{i\mu\nu}))\right>_{\rm dual}
\end{equation}
where ${\tilde t}^{(\rho)}$ denotes the matrix element of the
representation of $Z(G)$ that is induced by $\rho$.

%
\section{Some comments}
%

It should be pointed out that the dual Boltzmann weight $\hat f_\rho$
is strictly positive for the standard actions. However, the
coefficients $C(i)$ and $D(i)$ are not, so that their sign has to be
associated to the observable while their modulus can be part of the
probability density. 

The heat kernel action is defined by its dual Boltzmann weight $\hat
f_\rho = d_\rho\cdot\exp(-C_\rho/\beta)$ where $C_\rho$ is the
eigenvalue of the second order Casimir operator in the representation
$\rho$. The configurations of the dual path integral can therefore be
ordered according to their $\beta$-dependence. One just sorts them by
the value of the sum of the second order Casimir eigenvalues of the
representations that are attached to the plaquettes. The strong
coupling expansion of the generic spin network expectation value can
thus be read off from its dual expression.

Interesting future applications of the duality transformation include
not only the study of the topological degrees of freedom in connection
with an understanding of confinement, but also the development of
efficient algorithms for the dual model. Finally, the relation of the
dual form of LGT with models studied in the context of
non-perturbative quantum gravity suggests that both areas can benefit
from a combination of the available techniques, numerically and
analytically. 

\noindent
{\bf Acknowledgements}

R.O. is grateful to NATO for a Research Fellowship, H.P. would like to
thank Emmanuel College, Cambridge, for a Research Fellowship. Thanks
are also due to I.~Drummond, J.~Jers{\'a}k, A.~J.~Macfarlane,
S.~Majid and T.~Neuhaus for valuable discussions.

\end{document}